# Minibands in twisted bilayer graphene probed by magnetic focusing

A. I. Berdyugin[1], B. Tsim[1,2], P. Kumaravadivel[1,2], S. G. Xu[1,2], A. Ceferino[1,2], A. Knothe[2], R. Krishna Kumar[1,2], T. Taniguchi[3], K. Watanabe[3], A. K. Geim[1,2], I. V. Grigorieva[1,2,4], V. I. Fal'ko[1,2,4]

[1]School of Physics and Astronomy, University of Manchester, Manchester M13 9PL, UK.
[2]National Graphene Institute, University of Manchester, Manchester M13 9PL, UK.
[3]National Institute for Materials Science, 1-1 Namiki, Tsukuba 305-0044, Japan.
[4]Henry Royce Institute for Advanced Materials, Manchester, M13 9PL, UK.

**Magnetic fields force ballistic electrons injected from a narrow contact to move along skipping orbits and form caustics. This leads to pronounced resistance peaks at nearby voltage probes as electrons are effectively focused inside them, a phenomenon known as magnetic focusing. This can be used not only for the demonstration of ballistic transport but also to study the electronic structure of metals. Here we use magnetic focusing to probe narrow bands in graphene bilayers twisted at ~2°. Their minibands are found to support long-range ballistic transport limited at low temperatures by intrinsic electron-electron scattering. A voltage bias between the layers causes strong valley splitting and allows selective focusing for different valleys, which is of interest for using this degree of freedom in frequently-discussed valleytronics.**

Crystallographic alignment of atomically thin crystals stacked together in a van der Waals heterostructure is a powerful tool that enables fine tuning of their electronic spectra. For crystals with similar honeycomb lattices the spectra are modified by the presence of a long-range interference (moiré) pattern with a period $\lambda_s$ dependent on the twist angle θ between the layers (*1–18*), see Fig 1A. The additional spatial periodicity reduces the size of the Brillouin zone and introduces secondary Dirac points, as illustrated in Fig. 1B. So far, the most pronounced twist-engineered changes in the electronic properties of 2D crystals have been achieved in twisted bilayer graphene (TBG), where the twist at discrete 'magic' angles results in narrow bands, periodically modulated interlayer hybridisation and strong enhancement of electron correlations, leading to superconductivity and Mott insulator transitions (*6–8*). At larger θ, the TBG spectrum corresponds to a metal with several minibands at each K and K' valley in the Brillouin zone (Fig. 1B). Electronic properties of such a metal are expected to be quite different from the behaviour of Dirac electrons in monolayer or bilayer (aligned to Bernal stacking) graphene but so far remain largely unexplored. Here we use transverse focusing of electrons in a perpendicular magnetic field (TMF) (*12*, *19–23*) to probe the properties of moiré minibands in TBG and demonstrate an exceptionally high quality of the 'artificial metal' in TBG, as well as a possibility to use vertical displacement field, *D,* to break the valley degeneracy in the two constituent layers and selectively enhance transport in one of the valleys.

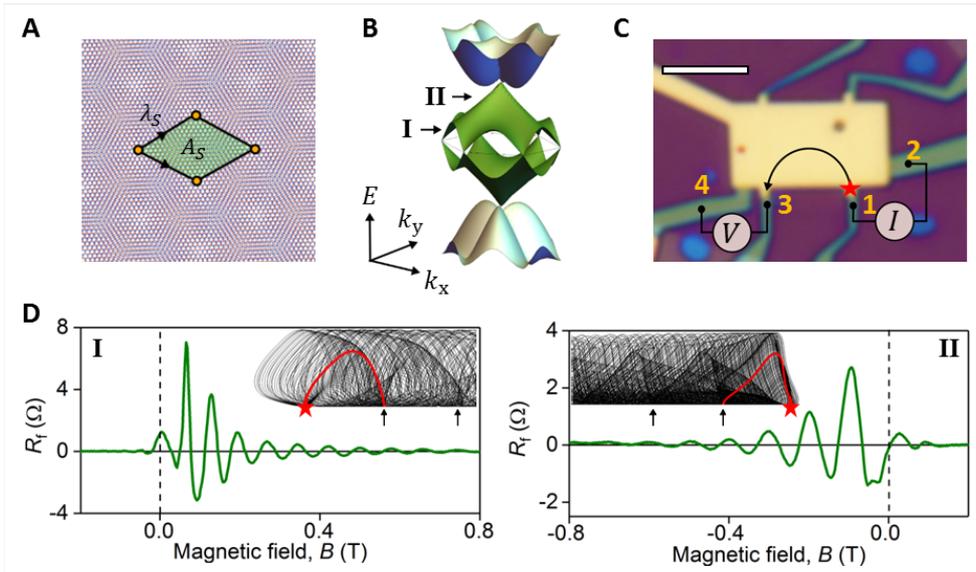

**Fig. 1. Moiré minibands and TMF measurements.** (**A**), *Schematics of the moiré superlattice induced by the twist of graphene layers. Here two graphene sheets are rotated by an angle θ relative to each other, which creates an additional spatial periodicity $\lambda_S = a/[2\sin(\theta/2)]$ (a is graphene's lattice constant) with the unit cell area of $A_S = \sqrt{3}/2\,\lambda_S^2$. (**B**) Band structure of TBG graphene in the K valley of the Brillouin zone calculated for the twist angle $\theta = 1.87°$ as discussed in (24), section 3. (**C**), Optical image of TBG device D1 with $\theta = 1.87°$. Scale bar corresponds to 4 µm. (**D**), Two examples of TMF signals measured in device D2 (D=0 V/nm) at 5K for the carrier density $3.7\times10^{12}$ cm$^{-2}$ (left) and $9.3\times10^{12}$ cm$^{-2}$ (right) at a distance 4.9 µm from the injector. The latter are close to the main and secondary neutrality points, respectively, as illustrated in panel (B). The insets are examples of focusing caustics near the main (left) and secondary (right) neutrality points. Arrows highlight the focal points for caustics, red star marks the current injection point and red lines show typical trajectories that extend from the injector to the first focal point.*

We studied two high-quality dual gated TBG devices encapsulated with ~30-50 nm thick hBN crystals: D1, with θ =1.87±0.01° (shown in Fig. 1C) and D2, with θ =2.60±0.01° (fig. S1A). The procedure used to determine θ is described in section 1 of Supplementary Materials (*24*). The devices were fabricated using standard dry-transfer (*25*, *26*) and tear-and-stack (*4*) techniques, see section 2 in (*24*) for details. To ensure a clean interface between the two graphene layers, special care was taken to avoid any contact between graphene and the polymer during the transfer (*24*). In transport measurements both devices showed similar behaviour, with low-temperature mobilities in excess of 400 000 cm$^2$V$^{-1}$s$^{-1}$ for carrier density $n\sim10^{12}$cm$^{-2}$. All data shown below were obtained at a constant displacement field, *D*, that was achieved by a simultaneous sweep of the top and bottom gates (*24*).

The high mobility for both devices enabled observation of TMF (*12*, *19–23*), which is a manifestation of ballistic motion of electrons and had been used to characterize the shape of Fermi surfaces both in 3D (*19*, *20*) and 2D (*12*, *21–23*) metals. To measure the effect of TMF in our TBG devices, we employed a nonlocal geometry illustrated in Fig. 1C, where narrow contacts 1 and 2 at one end of the device were used for current injection (driving current $I_{12}$) and contacts 3 and 4 at the other end were used to detect a voltage $V_{34}$. In the presence of a perpendicular magnetic field, electrons injected from contact 1 propagate along the device edges in skipping orbits and form a characteristic caustic pattern determined by the shape of the Fermi surface, as illustrated in the insets of Fig. 1D. Caustics are focused into equidistant focal points along the sample edge and the drift direction of the skipping orbits is determined by the sign of the magnetic field, such that electron-like and hole-like carriers propagate in opposite directions. As the positions of focal points vary with the magnetic field, whenever they coincide with the position of the voltage probe (contact 3 in Fig. 1C), one observes a focusing peak in the nonlocal resistance $R_f = V_{34}/I_{12}$. Fig. 1D gives two examples of the observed focusing peaks measured at different carrier densities.

Fig. 2A shows a typical dependence of $R_f$ on the carrier density and magnetic field at zero displacement field, $D=0$ V/nm. Here the appearance of an $R_f$ signal in a particular quadrant of the $B$-$n$ diagram reflects the sign of the cyclotron mass, while the change of the quadrant upon doping indicates an inversion of the electron dispersion (i.e., a change of sign of the mass from electrons to holes or vice versa). Accordingly, a fan-like pattern in the centre of Fig. 2A, which converges and changes direction at zero carrier density, indicates a neutrality point. Two additional, qualitatively similar, changes of the cyclotron mass appear at higher electron and hole densities, showing inverted fan-like patterns at higher energies. These indicate that the electron dispersion converges towards a new (secondary) neutrality point, such as shown in Fig. 1B. The crossover between these two regimes (at $n \approx 3\times10^{12}$ cm$^{-2}$ and $-3\times10^{12}$ cm$^{-2}$) must correspond to a van Hove singularity (vHS) in the moiré miniband spectrum. For a quantitative comparison, Fig. 2B presents the results of TMF modelling for device D1. Here we used the model from ref (3) to compute the electron spectrum shown in Fig. 1B [see section 3 in (24) for details] and to perform numerical analysis of caustics (12), where the contributions to $R_f$ from trajectories of electrons leaving the injection contact at different angles were weighted proportionally to $|\nabla_{\mathbf{k}}E|^{-1}$ [section 4 in (24)]. A good agreement between the experiment (Fig. 2A) and theory (Fig. 2B) suggests that the band structure of TBG is well described by the spectrum shown in Fig. 1B.

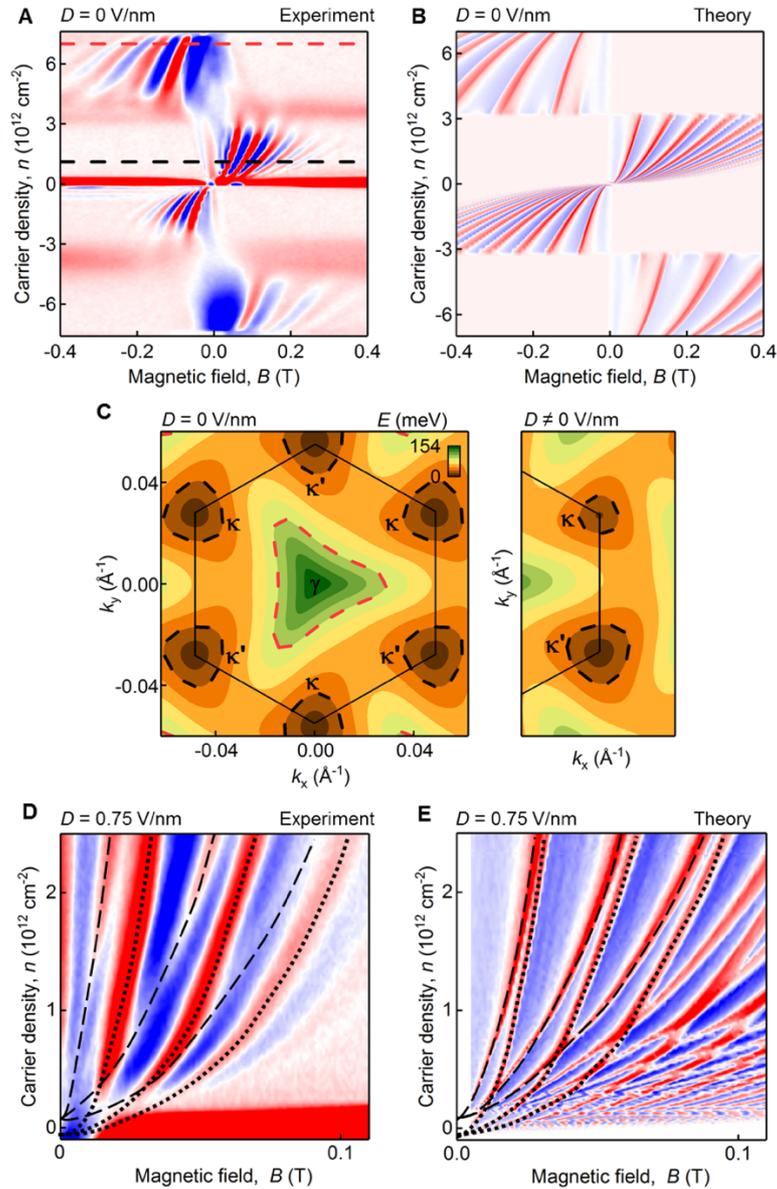

**Fig. 2. Transverse magnetic focusing map.** (**A**), *Focusing signal $R_f$ as a function of the magnetic field and carrier density measured at 2K for device D1 in zero displacement field, D=0 V/nm. Colour scale: blue to red ±3 Ω.* (**B**), *TMF map calculated from the energy spectrum shown in Fig. 1B using a numerical method described in section 4 of ref (24). The angle between the zigzag edge of one of the monolayers and the sample boundary is taken as 45˚ to avoid any spurious effects of crystallographic alignment. Importantly, as demonstrated in section 4 of ref (24), the calculated TMF map is only very weakly sensitive to the mutual orientation between graphene and the sample edge, confirming the generality of our results.* (**C**), *Contour plot of the first conduction miniband shown for the K valley of the Brillouin zone for zero (left) and non-zero (right) displacement fields. Black and red dashed lines outline the shape of the Fermi surfaces for carrier densities marked by black and red dashed lines in panel (A); the latter correspond to equivalent doping levels relative to the main (black) and secondary (red) neutrality points.* (**D**), *$R_f$ as a function of magnetic field and carrier density for device D2 measured at T=2K and D=0.75 V/nm at a distance of 8.5 µm from the injector. Dashed and dotted lines are guides to the eye, emphasising the first three focusing peaks from different valleys. Colour scale: blue to red ±0.2 Ω.* (**E**) *TMF map calculated numerically for device D2 in a displacement field (see sections 3, 4 and 5 in ref. (24) for details).*

It is noticeable that the fan-like patterns in Figs 2A and 2B - corresponding to the main and secondary neutrality points of the TBG superlattice (around zero carrier density and above the vHS, respectively) - have different periodicities. This difference is caused by different sizes of the Fermi surfaces at equivalent doping levels (black and red dashed lines in Fig. 2A), due to the degeneracy of the miniband dispersion at κ and κ'. The Fermi surface contours are shown in Fig. 2C by black dashed lines around κ and κ' points of the mini Brillouin zone (main neutrality point) and a red dashed line around the γ point (secondary neutrality point). Furthermore, our theoretical analysis suggests that the Fermi surfaces close to the γ point have a triangular shape (Fig. 2C) which can be traced to the strong interlayer hybridisation of those states. At the same time the Fermi surfaces around κ and κ' points (that coincide with the valley centres K of the top and bottom graphene layers) are almost isotropic, as in monolayer graphene, pointing towards weak interlayer hybridisation of these states.

The absence of appreciable interlayer coupling at κ and κ' can be used to disentangle the TMF contributions from different valleys. To this end, we employed a finite displacement field, up to D=0.75 V/nm (achievable without a risk of damaging our devices) which shifts the on-layer potential for electrons and, therefore, shifts the energies of the Dirac cones at κ and κ', as illustrated in Fig. 2C. Such layer-symmetry breaking lifts the degeneracy between κ and κ' and separates the motion of electrons from different valleys in a magnetic field, as they now have different sizes of cyclotron orbits. This generates two different magneto-oscillation frequencies of $R_f$ at low carrier densities, $|n|<10^{12}$ cm$^{-2}$, as clearly seen in Fig. 2D where separate focusing peaks appear for the electrons from each valley.

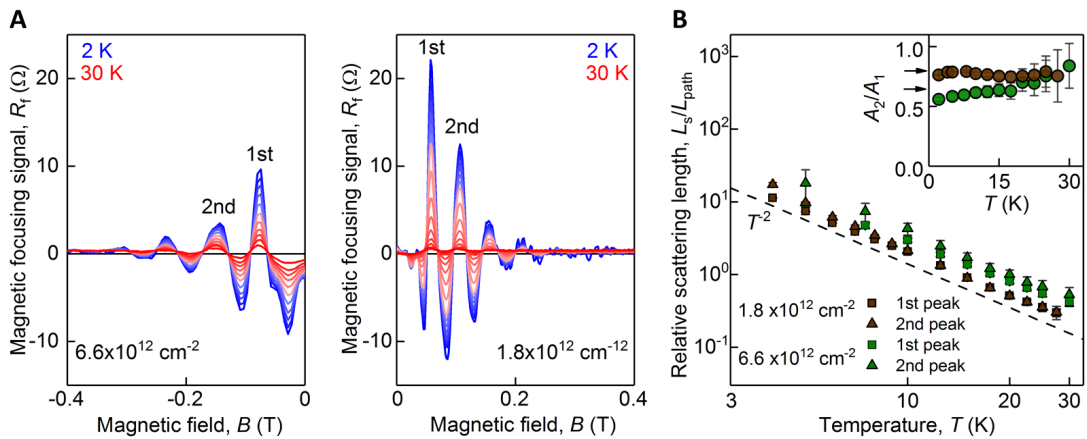

**Fig. 3. Temperature dependence of magnetic focusing.** (**A**), *Temperature dependence of the TMF signal measured at two characteristic carrier densities for device D1 (see legends). T was varied from 2K to 30K (blue to red).* (**B**), *T dependence of the relative scattering length (see text) extracted from experimental data for*

*consecutive focusing peaks. Dashed line shows $T^2$ dependence. The inset shows the ratio of the areas under the 1st and 2nd focusing peaks in (A) as a function of T. Arrows correspond to $A_2/A_1$= 0.8 and 0.65 (see text). Error bars indicate the accuracy of determining $A_2/A_1$; large errors at T>20K are due to the relatively large background signal as the focusing peaks become strongly suppressed.*

Further information about carrier dynamics in TBG can be obtained by studying the temperature dependence of TMF and its evolution for consecutive focusing peaks. In Fig. 3A we show how the amplitude of TMF oscillations depends on temperature $T$ in the range 2K < $T$ < 30K, in the vicinity of both main and secondary neutrality points. For quantitative analysis, we extract the relative scattering length as (*12*):

$$\frac{L_\mathrm{s}}{L_\mathrm{path}} = \left(\ln\left[\frac{A(T_\mathrm{base})}{A(T)}\right]\right)^{-1} \qquad (1)$$

where $L_\mathrm{path}$ is the length of trajectories extending from the injector to the first focal point as shown in Fig. 1D, and $A(T)$ and $A(T_\mathrm{base})$ the areas under the first focusing peak in Fig. 3A at $T$ and $T_\mathrm{base}$ =2K, respectively. The results are shown in Fig. 3B. The measured scattering lengths $L_\mathrm{s}$ for both carrier densities and all focusing peaks clearly follow a $T^{-2}$ scaling, which is different from the $T^{-1}$ dependence characteristic of phonon-dominated scattering (*23, 27*). Such scaling points towards the dominance of low-angle electron-electron scattering that was also found to be responsible for the TMF suppression in graphene/hBN superlattices (*12*). Furthermore, the ratio between the areas under the 2nd and 1st focusing peaks in Fig. 3A, $A_2/A_1$, characterizes the reflection of electrons at the sample boundary: The closer it is to one, the higher the probability for the incoming electrons to undergo specular reflection. In our experiment, electrons with energies near the main neutrality points ($n \approx$ 1.8×10$^{12}$ cm$^{-2}$, right panel in Fig. 3A) undergo almost specular reflection ($A_2/A_1 \approx$ 0.8), while reflection of the electrons with energies near the secondary neutrality point ($n \approx$ 6.6×10$^{12}$ cm$^{-2}$, left panel in Fig. 3A) is notably less specular ($A_2/A_1 \approx$ 0.65). This indicates a higher probability of diffusive scattering in the latter case, which is consistent with the greater sensitivity of the corresponding part of the miniband spectrum to inevitable perturbations of the moiré pattern near the sample edge. Indeed, due to little hybridization between the layers near κ and κ', the scattering of Dirac electrons should be little affected by the termination of superlattice periodicity near the edge, while its part near the secondary neutrality points should be affected significantly, promoting diffusive scattering.

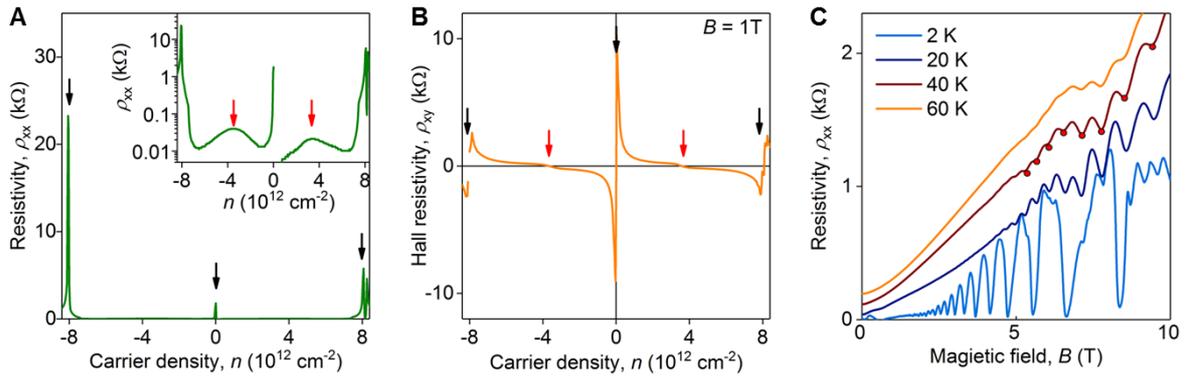

**Fig. 4. Bulk transport properties of TBG.** (**A**), *Resistivity as a function of carrier density measured at 5K for device D1. The inset shows the same data on a logarithmic scale.* (**B**), *Hall resistivity as a function of the carrier density for D1. Black arrows in (A) and (B) mark neutrality points and red arrows mark vHS.* (**C**), *Resistivity vs magnetic field measured at different temperatures for device D1 at n = 3.1x10$^{12}$ cm$^{-2}$. Red dots on a 40K curve highlight the positions of Brown-Zak oscillations.*

Finally, we note that the above observations of superlattice effects in TMF correlate well with the bulk transport properties of the same TBG samples studied using local geometry. The longitudinal and Hall resistivity shown in Figs 4A and 4B displays secondary neutrality points (indicated by black arrows) and vHSs (red arrows) at the same carrier densities as those inferred from the TMF experiments. Furthermore, the presence of a moiré superlattice in the studied TBG samples is seen from the presence of Brown-Zak oscillations (*13*, *14*) that dominate the magnetotransport above $T$~30K (Fig. 4C): While at low temperatures the magnetoresistance is dominated by Shubnikov-de Haas oscillations (see $T$ = 2K curve in Fig. 4C), these are rapidly suppressed as $T$ increases and give way to another 1/B-periodic oscillations, with period determined by the relation between the magnetic flux through the moiré supercell area, $A_S$, and the magnetic flux quantum $\phi_0 = h/e$, i.e., $BA_S = \phi_0/q$ (where $q$ is an integer).

To conclude, we have demonstrated that TBG supports ballistic propagation of electrons in multi-micrometer devices, with electron transport determined by the reconstruction of the energy spectrum in the presence of a long-period superlattice. This offers new opportunities to study fundamental phenomena, such as Bloch oscillations in moiré superlattices (*28–30*) and their use for, e.g., THz generation. Moreover, we have shown that the unique sensitivity of the TBG band structure to the displacement field allows selective manipulation of electrons from different valleys which can be implemented in the next generation of electronic devices based on the valley degree of freedom.

**Acknowledgements:** We thank J. R. Wallbank for helpful discussions. A. I. B., B. T., A. C. acknowledge support from Graphene NOWNANO Doctoral Training Center, R. K. K. was supported by an EPSRC fellowship award.

**Funding:** This work has been supported by EPSRC grants EP/S019367/1, EP/S030719/1, EP/N010345/1; EPSRC Doctoral Training Centre Graphene NOWNANO EP/L01548X/1; ERC Synergy Grant Hetero2D; Lloyd's Register Foundation Nanotechnology grant; European Graphene Flagship Project.

**Author contributions:** P. K. and S. G. X. fabricated devices. A. I. B. carried out electrical measurements with the help from R. K. K. The results were analyzed by A. I. B., B. T., V. I. F., I. V. G. Magnetic focusing simulations were done by B.T. with contributions from V. I. F., A. C. and A. K. T. T. and K. W. provided hBN crystals. A. I. B., B. T., I. V. G., V. I. F. wrote the manuscript with contributions from A. K. G., P. K. and S. G. X. All authors contributed to discussions.

**Competing interests:** The authors declare no competing interests.

**Data and materials availability:** All data needed to evaluate the conclusions in the paper are present in the paper and/or the Supplementary Materials. Additional data available from authors upon request.